\documentclass[12pt]{article}
\usepackage{epsfig,graphicx,bm,amsmath}
\newcommand{\be}{\begin{equation}}
\newcommand{\bea}{\begin{eqnarray}}
\newcommand{\eea}{\end{eqnarray}}
\newcommand{\ba}{\begin{array}}
\newcommand{\ea}{\end{array}}
\newcommand{\ee}{\end{equation}}

\expandafter\ifx\csname mathbbm\endcsname\relax

\else

\fi \textheight 22cm \textwidth 15cm \topmargin 1mm
\def\one{(\gamma_1/(\gamma_1-\gamma_2))}
\def\two{(\gamma_2/(\gamma_2-\gamma_1))}
\begin{document}

\begin{titlepage}
\hfill
\begin{flushright}
{IFT-P.040/2006}\\
\end{flushright}

\vspace*{10mm}
\begin{center}
{\Large {\bf Non-Supersymmetric Attractors in BI black holes}\\ }
\vspace*{15mm} \vspace*{1mm} {B. Chandrasekhar$^{a}$
\footnote{chandra@ift.unesp.br},   H. Yavartanoo$^{b}$
\footnote{yavar@phya.snu.ac.kr}, and  S. Yun$^c$ }
\footnote{sanhan1@phya.snu.ac.kr}\\
\vspace*{1cm}
{$^a$ Instituto de F\'{i}sica Te\'{o}rica, Universidade Estadual Paulista, \\
Rua Pamplona, 145, 01405-900, S\~{a}o Paulo, SP, BRASIL
\\ \vspace{3mm}
$^b$ Center for Theoretical Physics and BK-21 Frontier Physics Division,
Seoul National University, Seoul 151-747 KOREA \\ \vspace{3mm}$^c$ School of Physics and Astronomy, Seoul National University\\
Seoul 151-747, KOREA \\
\vspace*{3mm}
}
\end{center}
\begin{abstract}
We study attractor mechanism in extremal black holes of Einstein-Born-Infeld
theories in four dimensions. We look for solutions which are regular near the horizon and show that they exist and enjoy the attractor behavior. The attractor point is determined by extremization of the effective potential at the horizon. This analysis includes the backreaction and supports the validity of non-supersymmetric attractors in the presence of higher derivative interactions in the gauge field part.\end{abstract}
\end{titlepage}
\section{Introduction} 

The study of attractor mechanism \cite{9508072}-\cite{9702103} in extremal black holes of general theories of gravity and
string theory has drawn a lot of attention recently \cite{0405146}-\cite{Astefanesei:2007bf} . This, in part, is due to the realization
that the concept of attractor mechanism is rather general and goes beyond the original motivation,
where supersymmetry was the key ingredient. This mechanism can be used to study the properties of
extremal black holes in supersymmetric theories, which do not respect any supersymmetry, or those in non-supersymmetric theories. In all these cases,
the statement of attractor mechanism is that, in a generic
situation, the near horizon geometry and the black hole entropy turn out to be completely independent
of the asymptotic behavior of scalar fields of the theory and depend only on certain conserved quantities, like mass, charge and angular momentum.
In this context, the entropy function formalism \cite{0506177} has proved to be a very useful tool for calculating
the entropy of extremal black holes in a general theory of gravity, with any set of higher derivative
terms and in higher dimensions. This formalism is based on the fact that, knowing the near horizon
symmetries of the black hole is enough to generate the entropy through the use of Wald's entropy formula, and attractor equations are essentially some linear combinations of the equations of motion of all the fields of the theory.

\vskip 0.3cm

The concept of attractor mechanism is extremely useful in calculating the
entropy of extremal black holes. It also turns out to be helpful in seeking to comprehend the structure of higher
derivative terms in a general theory of gravity \cite{0007195}-\cite{Sahoo:2006pm}. String theory at low energies gives rise to a variety of higher derivative terms. Understanding the structure of these higher derivative terms is of paramount importance, not just for the case of black holes, but also because they hold a lot of information about the unitarity and renormalizability properties of the theory in question.
With the use of the tools provided by attractor mechanism and entropy function formalism, interesting aspects of Lovelock terms, Chern-Simons terms, Born-Infeld terms etc., can be studied \cite{0511306}-\cite{0604106}.  Although, most of the issues
have been studied keeping in mind that the black holes cannot have any hair, it is also important to study the structure of higher derivative terms in the case when the black hole has hair. One of the key reasons is that, in non-supersymmetric theories, the assumption of a flat potential for the scalar fields can be erroneous, in which case one is left with  scalar hair. This is also
important from the point of view of the need to introduce a small amount of non-extremality, in
certain situations involving higher derivative terms~\cite{0611143}.

\vskip 0.3cm

Although considerable progress has been made in understanding the physics of  attractor
mechanism, there are a number of issues which need to be addressed, especially, in the absence of
supersymmetry. For instance, in the absence of supersymmetry, though non-supersymmetric
attractor mechanism can be used to calculate the entropy of black holes, the existence of
a full black hole solution interpolating between the Bertotti-Robinson geometry and the
asymptotically flat space is not guaranteed. This issue becomes more difficult to deal with when there are higher derivative terms in addition. Thus, it becomes important to formulate
and study non-supersymmetric attractor mechanism when there are different kinds of higher
derivative terms following from the low energy limit of string theory, in the Einstein action. Another issue is the stability of solutions of
non-supersymmetric attractor equations. One of the methods to address the problem of
stability of attractor points is to follow~\cite{9702103,0507096} and check on a case
by case basis.

\vskip 0.3cm

In this note, we study attractor mechanism in a general Einstein-Born-Infeld theory of gravity
coupled to moduli fields. Born-Infeld terms are known to arise in the low energy limit of
a configuration where gauge fields are coupled to open bosonic or superstrings. In fact, the
low energy theory on the world-volume of a $D$-brane is governed by a Born-Infeld action.
The importance of Born-Infeld terms in the context of extremal black holes and their
connection with elementary string states was stressed in~\cite{9506035}. It was argued that
virtual black holes going around closed loops can give rise to Born-Infeld type
corrections to extremal black hole configurations with non-trivial dilaton profiles.
On the otherhand, Einstein-Born-Infeld black holes in
presence of string generated low energy fields have been studied, for
example in~\cite{Wiltshire:1988uq}-\cite{0101083}. Thus, it is important to study if
the attractor mechanism works in the case of extremal black holes in Einstein-Born-Infeld theory.
Furthermore, if the mechanism works, then, the entropy function formalism can be used
to calculate the entropy in this case~\cite{Chandrasekhar:2006zw}.

\vskip 0.3cm

The rest of this paper is organized as follows. In section 2, we start by recollecting relevant features of attractor mechanism needed for our purposes in the case of  Einstein-Maxwell theory coupled to scalar field and discuss the possible attractor solutions in a general theory of gravity coupled to gauge fields and scalars.  Section 3 is devoted to studying attractor mechanism in Einstein-Born-Infeld theory coupled to a scalar field.  In section 4 we present a general perturbative analysis to show the presence of
attractor mechanism in this theory. Our conclusions are summarized in section 5.

\section{Non-Supersymmetric Attractors: General features } 

Let us start with a few relevant aspects of non-supersymmetric attractors which are needed
for our purposes. We consider the class of following gravity theories coupled to $U(1)$ gauge fields and scalar fields as in \cite{0507096}:
\be
\label{EHaction}
  S=\frac{1}{\kappa^{2}}\int d^{4}x\sqrt{-g}\left(R-2(\partial\phi_i)^{2}-
  f_{ab}(\phi_i)F^a_{\mu \nu} F^{b \ \mu \nu} - \frac{1}{2\sqrt{-g}}{\tilde f}_{ab}(\phi_i) F^a_{\mu \nu}
  F^b_{\rho \sigma} \epsilon^{\mu \nu \rho \sigma} \right) \, ,
 \ee
where $F^a_{\mu\nu}, \; a=0,...N$ are gauge fields and $\phi^i, \; i=1,...n$ are scalar fields. The scalar-dependent couplings of gauge fields are motivated from analogy with
the supersymmetric theories. Any additional potential term for the scalar fields will
lead to a breakdown of attractor mechanism in asymptotically flat spaces.
Rest of the notations are as in \cite{0507096}.

A static spherically symmetric ansatz is:
\bea
\label{metric}
ds^2&=&-\alpha(r)^2dt^2+\alpha(r)^{-2}dr^{2}+\beta(r)^2d\Omega^2 \, .
\eea
On the other hand, the Bianchi identity and equations of motion of gauge fields can be solved by taking the gauge field strengths to be of the form:
\be \label{gaugefield}
F^a=f^{ab}(\phi_i)(Q_{eb}-\tilde{f}_{bc}Q^c_m)\frac{1}{\beta^2}dt\wedge dr + Q^a_m \sin\theta d\theta \wedge d\varphi,
\ee
where $Q^a_m$ and $Q_{ea}$ are constants that determine the magnetic and electric charges carried by the gauge fields $F^a$, and $f^{ab}$ is inverse of $f_{ab}$.

For a brief recap of non-supersymmetric attractor mechanism, it is enough to
concentrate only on the equations of motion of scalar fields:
\be
\partial_r(2\alpha^2\beta^2\partial_r\phi_i) = \frac{1}{\beta^2} \, \partial_i V_{eff}, \label{eqphi} \\
\ee
with the effective potential given by,
\be \label{effective}
V_{eff}(\phi_i) = f^{ab}(Q_{ea}-\tilde{f}_{ac}Q^c_{m})(Q_{eb}-\tilde{f}_{bd}Q^d_{m}) + f_{ab}Q^a_{m}Q^b_m.
\ee
Non-supersymmetric attractor equations can be derived
from $\partial_i V_{eff}(\phi_{i0})=0$, which also determines the attractor values
of scalar fields in terms of the fixed charges of the extremal black hole. This effective potential
can in fact be shown to be equivalent to Sen's entropy function prescription, as discussed in~/cite{0601016}.

\vskip 0.3cm

For matching the microscopic and macroscopic results for the entropy, it is important to consider
the impact of higher derivative terms. For a particular set of curvature squared terms in ${\cal N} =2$
supergravity, the corrections to entropy can be calculated~\cite{0009234}. For non-supersymmetric
extremal black holes, the effect of higher derivative corrections can also be calculated as
in~\cite{0508042}. As discussed in \cite{Chandrasekhar:2006kx}, using the set up described in this
section, non-supersymmetric attractor mechanism can be shown to be present when one
includes a certain set of higher derivative terms coming from the gravity side in the Einstein-Maxwell
action. It was also argued in \cite{Chandrasekhar:2006kx} that, in the presence of general
$R^2$ terms in the action, the effective potential gets modified by additional terms, and was
in fact called as $W_{eff}$. The scalar field equation of motion remains as in (\ref{eff}), with $V_{eff}$ replaced by $W_{eff}$.

\vskip 0.3cm

Before proceeding, it should be mentioned that $W_{eff}$ will in general depend on $r$. However, near
the horizon all the quantities are independent of $r$. In this special situation, the
$r$ dependence in $W_{eff}$ drops out. As a result, the horizon radius computed from
$W_{eff}$ will also be a constant, but modified by higher derivative terms. For instance,
let us note down the general form of the scalar field equation near the horizon,
in the presence of Gauss-Bonnet terms:
\be
\left(\alpha^2 \beta^2\phi'\right)' = \frac{1}{2\beta^2}\frac{d W_{eff}}{d\phi} \, ,
\ee
where,
\be
W_{eff}(\phi) = V_{eff}(\phi) + 4\, a\, G(\phi) \, ,
\ee
and there is no $r$ dependence. The additional term $4\, a\, G(\phi)$ also modify the entropy of the black hole via Wald's entropy formula. This is in parallel to the analysis in Sen's entropy function formalism, where the addition of Gauss-Bonnet term gives rise to a finite area to the horizon and hence to the entropy of small black holes in heterotic string theory.

\vskip 0.3cm

In the following sections, we show that the above analysis of~\cite{0507096,Chandrasekhar:2006kx} can
be extended to include higher derivative interactions in the gauge field part, in particular to the
case of Born-Infeld terms.
Black hole solutions in Einstein-Born-Infeld theories have been studied quite a lot in
literature. It is known that one can have particle-like and BIon solutions in these theories.
However, finding explicit black holes solutions in the presence of scalar couplings in the Einstein-Born-Infeld action is non-trivial. In four dimensions, when looking for asymptotically
flat solutions in these theories, it is reasonable to assume that the near horizon geometry of
these black holes preserve the symmetries of $AdS_2 \times S^2$.
In order to understand the effect of higher order Born-Infeld corrections to the entropy
of extremal black holes, an entropy function analysis of small black holes in heterotic string
theory was presented in \cite{Chandrasekhar:2006zw}. However, it is important to check if the
attractor mechanism works when considering the full black hole solution. As in \cite{0507096,Chandrasekhar:2006kx}, in this work, we carry out a perturbative analysis
to show that the moduli fields take fixed values as they reach the horizon and that a double horizon Einstein-Born-Infeld black hole continues to exist.
We show that the attractor mechanism works in the case of Born-Infeld black holes.
In effect, we show that, once one obtains critical
values of the effective potential and ensures that $\partial_i \partial_j V_{eff}(\phi) > 0$,
the perturbative analysis signifies that there is always a solution of equations of motion where
the scalar fields get attracted to fixed points, which remain stable.

\section{\bf Non-Supersymmetric Attractors in Einstein-\\Born-Infeld theories}

Non-supersymmetric attractor mechanism in Einstein-Born-Infeld theories can
be studied using the entropy function formalism~\cite{Chandrasekhar:2006zw}.
However, to see that the moduli
indeed get attracted to fixed points near the horizon, one has to use
the formalism for non-supersymmetric attractor mechanism reviewed in the
previous section, which
makes explicit use of the general solutions and equations of
motion for two~\cite{0507096} and higher derivative~\cite{Chandrasekhar:2006kx} gravities.
\vskip 0.2cm

In this section, we follow the analysis outlined
in the previous section. Using a perturbative approach to study the corrections
to the scalar fields and taking the backreaction corrections into the
metric, it is possible to show that the scalar fields are indeed drawn
to their fixed values at the horizon. Here, the requirements are the
existence of a {\it double degenerate horizon solution}. We now concentrate
on the analysis using equations of motion explicitly and study the attractor mechanism
in the case of Einstein-Born-Infeld black holes coupled to moduli
fields.

\subsection{Born-Infeld theory}
Born-Infeld theory is one of the most important generalizations to non-linear electrodynamics. 
It was proposed to obtain a finite self-energy of the electron in arbitrary dimensions as follows:
\be
{\cal L}_{BI} = 4b \left\{\sqrt{-det\eta_{\mu\nu}}  -  \sqrt{-det(\eta_{\mu\nu} + \frac{1}{\sqrt{b}}F_{\mu\nu})} \right\}.
\ee
where $\eta_{\mu\nu}$ and $F_{\mu\nu}$ represent the Minkowski metric and the electromagnetic field strength tensor, respectively, and $b$ is a parameter characteristic of the Born-Infeld dynamics, and measures the nonlinearity of the theory. Born-Infeld theory has received attention again for it plays a significant role in string theory. It arises naturally in open superstrings and in D-branes~\cite{plb163}-\cite{9908105}. In open superstring theory, loop calculations lead to the above Lagrangian with $b=(2 \pi \alpha')^{-2}$. It has also been observed that the Born-Infeld action arises as an effective action governing the dynamics of vector fields on D-branes.

In four-dimensional spacetime, the Lagrangian can be expanded out to be:
\be
{\cal L}_{BI} = 4b \left\{1  -  \left[1 +
\frac{1}{2b}F_{\mu\nu}F^{\mu\nu}  - \frac{1}
{16b^2}(F_{\mu\nu}\star F^{\mu\nu})^2  \right]^\frac{1}{2} \right\}
\ee
where $\star F^{\mu\nu}$ denotes the dual tensor, $\star F^{\mu\nu} = \frac{1}{2} \epsilon^{\mu\nu\rho\sigma} F_{\rho\sigma}$. The above Lagrangian reduces to the usual Maxwell one in the weak field limit. In the open superstring theory where the dilaton field is expressed by $\phi$, the Lagrangian is modified to (See \cite{Callan:1988LNY,9506035,0007228}):
\be \label{axon}
{\cal L}_{BI} = 4b e^{2\phi} \left\{\sqrt{-det g_{\mu\nu}}  - \sqrt{- det (g_{\mu\nu} + \frac{e^{-2\phi}}{\sqrt{b}}}F_{\mu\nu})  \right\}
\ee

\subsection{BI Attractor Equations}

For the purpose of studying non-supersymmetric 
attractor mechanism in Einstein-Born-Infeld black holes coupled to moduli
fields, we start from the following action with general couplings:
\be \label{last}
S=\frac{1}{16\,\pi}\int d^{4}x\sqrt{-g}\left(R_g
- 2\,(\partial\phi)^2 + 
{\cal L}_{BI} \right)  \, ,
\ee
where,\\
\be
{\cal L}_{BI} = 4b f(\phi)\left\{1  -  \left[1 +
\frac{f^{-2}(\phi)}{2b}F^{2}  - \frac{f^{-4}(\phi)}
{16b^2}(F\star F)^2  \right]^\frac{1}{2} \right\}.
\ee
where $F_{\mu \nu}$ denotes the field
strength and  $f(\phi)$ stands for a dilaton like coupling of the scalar field $\phi$. Various interesting BI solutions ensuing from the action were
studied in~\cite{0004071,0101083}. We have further tried to ensure that
the action in eqn. (\ref{last}), reduces to the action considered in~\cite{0507096} in the 
limit $b\rightarrow \infty$. The case considered in~\cite{0507096} is more general and 
from (\ref{axon}), in the following, we specialize to the case
$f(\phi)=e^{2\gamma \phi}$ where $\gamma$ characterizes the strength of dilaton field. 
It is one for string theory. We keep this parameter arbitrary, so that even more general theories of
gravity, apart from the ones descending from string theory could also be considered.
In the present case, we restrict ourselves to the case
of a single gauge field (excepting the example in section 3.3) 
and scalar field, and with the black hole carrying dyonic charges. 
It is important not to have a potential for the scalar fields, so as to allow for the
existence of a moduli space to vary.  In the absence of any moduli fields,
Einstein-Born-Infeld black holes have been constructed
in \cite{Wiltshire:1988uq}. In what follows, we shall be interested in
asymptotically flat spacetime solutions, although, the generalization to
include a cosmological constant should also be possible. In fact, it
might be interesting to include a cosmological constant~\cite{Dey:2004yt}
in view of the results in \cite{Chandrasekhar:2006zw}.

\vskip 0.2cm

Now, one makes an ansatz for a static spherically symmetric metric
which must satisfy the field equations following from the
Einstein-Born-Infeld action in eqn. (\ref{last}).  It should be
mentioned, that although we are working with a system of gauge fields
coupled to scalar fields, to lowest order,
we are looking for the solutions of the equations of motion only for constant
values of the moduli fields. The Birkhoff's theorem holds in this case and
we may assume the solution to be static and spherically symmetric,
to be of the form:
\bea
ds^{2} &=& -\alpha(r)^{2}\,dt^{2}\,+\;\frac{dr^{2}}{\alpha(r)^{2}}\;+\;
\beta(r)^{2}\;d\Omega^{2}_2 \, , \crcr
F &=& F_{tr} \; dt \wedge dr + F_{\theta\phi}\;d\theta \wedge d\phi.
\eea
The induction tensor $G_{\mu\nu}$ is defined by
\be
G^{\mu\nu} = -\frac{1}{2} \frac{\partial L}{\partial F_{\mu\nu}} \, .
\ee
The Maxwell equations and Bianchi identity are
\be
dG=0 \,\, , \hspace{2cm} \, dF=0 .
\ee
The above two equations give us the following solutions:
\be
F_{tr}= \frac{Q_e e^{2\gamma\phi}}{\beta^2 \sqrt{1+\frac{Q^{2}_e+Q^{2}_me^{-4\gamma\phi}}{b\beta^4}}} \, , \hspace{10mm} F_{\theta\phi}= Q_m\sin\theta \, \, .
\ee
Although, for simplicity, we consider the case of a single scalar field and a gauge field, the generalization
to many scalar fields is straightforward.
The equations of motion and the Hamiltonian constraint derived from the action $S$ with the above solution
for gauge fields and the metric ansatz turn out to be:
\bea
\label{eqn1}
-1 + \alpha^2\beta'^2 + \frac{{\alpha^2}'{\beta^2}' }{2}- \alpha^2\beta^2 (\partial_r \phi)^2 + \frac{1}{\beta^2}V_{eff} &=& 0 \, , \\ \cr\cr
\label{ab}
\alpha'^2 + \alpha \alpha'' + \frac{2\alpha\alpha'\beta'}{\beta} - 2 b  e^{2\gamma\phi}\left(1-\frac{1}{\sqrt{1+\frac{Q_e^2+Q_m^2e^{-4\gamma\phi}}{b\beta^4}}} \right) &=& 0  \, ,\\ \cr\cr
\label{dil}
\partial_{r}(2\alpha^{2}\,\beta^{2}\partial_{r}\phi)
- \frac{\partial_{\phi}V_{eff}}{\,{\beta^2}} &=& 0 \, , \\ \cr
\label{eqn4}
(\partial_r \phi)^2  + \frac{\beta''}{\beta}&=& 0 \, ,
\eea
where $V_{eff}$ plays the role of an `effective
potential' for the scalar fields.
A difference with \cite{0507096}
is that $V_{eff}$ in this case, is a function of $r$, as
seen below:
\be
\label{eff}
V_{eff} = 2b\,\beta^4 e^{2\gamma\phi}\left(
{\sqrt{1+\frac{Q_e^2+Q_m^2e^{-4\gamma\phi}}{b\beta^4}}} -1 \right) \,\, .
\ee
However, as discussed in~\cite{0601016}, it is possible to treat $r$ as just
a parameter near the horizon. Extremizing the effective potential
gives the fixed values taken by the moduli at the horizon.

\subsection{Exact Solution}

Let us study some exact solutions, which in the $b\rightarrow \infty$ limit give the solutions
considered in~\cite{Gibbons:1987ps,0507096}. For the time being, let us consider only the
electrically charged case, i.e., $Q_m=0$. For instance, if we have two gauge fields and a single
scalar field, then, the effective potential turns out to be:
\be
V_{eff} = 2b\,\beta^4 S^{-\gamma_1}\left(  {\sqrt{1+\frac{Q_1^2}{b\beta^4}}} -1 \right)
+ 2b\,\beta^4 S^{-\gamma_2}\left(  {\sqrt{1+\frac{Q_2^2}{b\beta^4}}} -1 \right) \, ,
\ee
where $S = e^{-2\phi_0}$ is the notation near the horizon. Extremizing the effective potential, the near horizon value of the scalar field gets fixed at:
\be\label{Sexact}
S = \left( \frac{\gamma_2}{\gamma_1}\, \frac{(  {\sqrt{1+\frac{Q_2^2}{b\beta^4}}} -1)}{(  {1- \sqrt{1+\frac{Q_1^2}{b\beta^4}}})}  \,\right)^{\frac{1}{\gamma_2-\gamma_1}} \, .
\ee
The second derivative of the effective potential is,
\be
\partial^2_{\phi}V_{eff} = S^{\gamma_2-2}\,(  {\sqrt{1+\frac{Q_2^2}{b\beta^4}}} -1) \,(\gamma_2^2-\gamma_1\gamma_2) ,
\ee
which is positive if $\gamma_1$ and $\gamma_2$ are of opposite sign.  The critical value
of the scalar field in eqn. (\ref{Sexact}) is also independent of the asymptotic value
of the moduli field. Area of the event horizon is,
\bea
{\rm Area } &=& 4\pi \beta_H^2 =  4 \pi\,V_{eff}(\phi_0)\, \, , \nonumber \\
&=& 8\pi\,b\,r_H^4\,\eta\,\left(  {\sqrt{1+\frac{Q_2^2}{br_H^4}}} -1 \right)^{\one}\,
\left(  {\sqrt{1+\frac{Q_1^2}{br_H^4}}} -1 \right)^{\two}\, \, ,
\eea
where, $r_H$ is the radius of the horizon and,
\be
\eta = \left(-\frac{\gamma_2}{\gamma_1}\right)^{\one} + \left(-\frac{\gamma_2}{\gamma_1}\right)^{(\gamma_2/(\gamma_1-\gamma_2))} \, \, .
\ee
In the case, $\gamma_1 = - \gamma_2$, we have,
\be
\frac{1}{4} {\rm Area} = 4\pi b\,r_H^4\,\left|\left(  {\sqrt{1+\frac{Q_2^2}{br_H^4}}} -1 \right)^{1/2}\right|\,\left|\left(  {\sqrt{1+\frac{Q_1^2}{br_H^4}}} -1 \right)^{1/2}\right| \, .
\ee

The radius of the horizon can be found as follows. Assuming a double horizon
solution to the equations of motion (\ref{ab})-(\ref{eqn4}) as,
\be
\alpha_0(r) = \alpha_H\left(1- \frac{r_H}{r} \right), \qquad \beta_0(r) = r ,
\ee
one can use the hamiltonian constraint in eqn. (\ref{eqn1}) to obtain,
\be
\beta_H^2 = V_{eff}(\phi_0) \, \, ,
\ee
where $\phi_0$ is the critical point of the effective potential. Solving this
equation for the special case of $\gamma_2 = -\gamma_1$ and $Q_1 = Q_2 = Q$, one
gets:
\be
r_H^2 = \beta_H^2 = 2 \left( Q^2 - \frac{1}{16b}\right) \, .
\ee
This result for the radius of the horizon is similar to~\cite{9506035}, after
some redefinitions, although in~\cite{9506035}, there were no moduli fields. Thus, we have,
\be
\frac{1}{4} {\rm Area} = 2 \pi \left( Q^2 - \frac{1}{16b}\right) \, \,
\ee

\section{Perturbative Analysis}

\vskip 0.2cm

It is well known that the equations of motion (\ref{eqn1})-(\ref{eqn4}),
admit $AdS_2 \times S^2$ as a solution
in the case of constant moduli. However we wish to address the attractor behavior
considering double horizon black hole solutions, which are asymptotically flat.
Thus, we start with  an extremal black hole solution in this theory, obtained
by setting the scalar fields at their critical values of the effective potential.
Then, as one varies the values of scalar fields at asymptotic infinity, we show
that the double horizon nature of black holes continues to exist. Further, the critical
values of the scalar fields remain stable, as the asymptotic values of these
moduli fields are somewhat different from attractor values.

\vskip 0.2cm

In view of the fact that the four
equations governing $(\alpha(r),\beta(r),\phi(r))$ are a set of highly
complicated coupled differential equations of order four, we follow
the Frobenius method to solve these equations as in~\cite{Chandrasekhar:2006kx}.
We call these four sets of equations of motion $EqA,EqB,Eq\Phi,EqC$.
As a variable of expansion we define $x \equiv (1-\frac{r_H}{r})$,
ranging from 0 to 1 to cover $r\geq r_H$ completely. Requiring that the solution:
(a) be extremal:  meaning that we have a double degenerate horizon as,  $\alpha^2(r)=(r-r_H)^2\tilde\alpha^2(r)$, with $\tilde\alpha^2(r)$
being analytic at the horizon $r=r_H$, (b) be asymptotically flat:
meaning that the black hole geometry tends to be flat and moduli fields take arbitrary values
at asymptotic infinity and
(c) be regular at the horizon\footnote{Just like the cases studied in references \cite{0507096,Chandrasekhar:2006kx} there is a solution where the scalar blows up at the horizon. In the
supersymmetric case, the well behaved solution is automatically choosen.},
the most general Frobenius expansions of
$\alpha(r)$,$\;\beta(r)$ and $\phi(r)$ take the form:
\bea\label{aexpan}
\alpha^2(r)&=&\; \alpha_H^2 x^2 (1+ \;\sum_{n=1}^{\infty}a_{n} x^{\lambda_1 n})
,  \\
\label{bexpan}
\beta(r)&=&\;  r (1 +\sum_{n=1}^{\infty} b_{n}x^{\lambda_2 n}) ,  \\
\label{phiexpan}
\phi(r)&=&\; \phi_0+\;\sum_{n=1}^{\infty}\phi_{n} x^{\lambda_3 n},
\eea
with $\lambda_i > 0$.
\vskip 0.2cm

When $V(\phi)$ of (\ref{eff}) is of pure magnetic (electric) type, the case given in (\ref{eff}) does not have an extremum for any finite value of $\phi$.
To have an extremum with electric or
magnetic fields and not both, one needs at least two gauge fields. Here we consider a dyonic case whence both electric and magnetic charges are non-zero.

\begin{flushleft}
\underline{{\bf Zeroth order results}}
\end{flushleft}
At zeroth order perturbation we start with a double horizon black hole solution as follows,
\bea\label{v0}
\phi(r) = \phi_0, \;\;\;\;\;\;\ \beta(r) = r ,\;\;\;\;\;\;\;\; \alpha(r) = \alpha_H (1-\frac{r_H}{r})\, \, ,
\eea
where, for given electric and magnetic charges, $\phi_0$, $\alpha_H$ and $r_H$, can be found from the following equations in terms of these charges,
\bea
e^{4\gamma\phi_0} &=& \frac{Q_m^2}{Q_e^2} -\frac{1}{4bQ_e^2} \, , \\ \cr
r_H^4 &=& 4 Q_e^2 Q_m^2 -\frac{Q_e^2}{b} \, , \\ \cr
\label{v1}
 \alpha_H^2 &=& 1- \frac{1}{4 b Q_m^2} .
\eea
We should mention, that from the above equations we find a lower bound for the value of magnetic charge to be $4b Q_m^2 \geq 1$. This bound relaxes in the limit $b\rightarrow \infty$, where, the Born-Infeld theory reduces to the Maxwell theory. In this limit $\phi_0$ and $r_H$ approach  values that one can find in Einstein-Maxwell-Dilaton theory~\cite{0507096}. In this case, a Reissner-Nordstrom  black hole with constant scalars, is an exact solution of the equations of motion.
\vskip 2mm

Notice that the equations (\ref{v0}-\ref{v1}), together, determine both the attractor value of the moduli field and the horizon radius in terms of the charges and the parameters of the action. In fact, both the above results are quite useful. Due to (\ref{v0}), the Bekenstein-Hawking entropy of the solution is given by the value of the $V_{eff}(\phi_0)$, up to a numerical prefactor.

\vskip 0.2cm

This, in fact fixes $\phi_0 $ at its extremum point.
From (\ref{phiexpan}), $\phi_0 = \phi(r_H)$ and so the value of
the moduli field is fixed at the horizon, regardless of any other
information. To complete the proof of the attractor behavior, one has to be able to show
that the four sets of equations of motion, denoting a coupled system of differential
equations, admit the expansions (\ref{aexpan}), (\ref{bexpan}) and
(\ref{phiexpan}). Also, one should see that there are
solutions to all orders in the expansion parameter $x$, with arbitrary
values taken by scalar fields at asymptotic infinity, where their
value at the horizon is fixed to be $\phi_0$.
The existence of a complete set of solutions with desired boundary conditions (considering the
fact that we have coupled non-linear differential equations) in the present case is very
interesting. Moreover, it is easy to show that, in our theory, there is no asymptotically flat solution
with everywhere constant moduli.

\begin{flushleft}
\underline{{\bf First order results}}
\end{flushleft}
To start with the first order perturbation theory, we write,
\be
\delta \phi \equiv \phi-\phi_0,
\ee
where, we keep $\delta \phi $ as a small parameter in perturbation theory. From the
equation of motion of the scalar field, we find,
\be
\label{phi}
\delta \phi = \phi_1 (1-\frac{r_H}{r})^k \, \, .
\ee
where,
\be \label{k}
k= {\frac{1}{2}(-1+\sqrt{1+8\gamma^2})}
\ee
Here, $\phi_1$ is an undetermined constant of integration. Since, we are considering
the case where $k>0$, in the asymptotic region $r \rightarrow \infty$,
we have $\delta \phi \rightarrow \phi_1$, which means that, there is a moduli space where the scalar fields can
take arbitrary values, since, $\phi_1$ can take arbitrary values. However, near the horizon, $\delta \phi$ vanishes, as seen from eq. (\ref{phi}), i.e., the value of the scalar field remains fixed at $\phi_0$ regardless of its asymptotic value. This shows that the attractor mechanism works
to first order in perturbation theory.

\vskip 0.2cm

In comparison to the Einstein-Maxwell theory where a Reissner-Nordstrom black hole case was considered in~\cite{0507096}, here we have a correction to the components of the metric at the first order in perturbation theory. At this order $\beta(r)$ does not get any correction, while $\alpha(r)$ receives corrections as follows:
\be
\alpha_1(r)= \alpha_H^2 a_1 (1-\frac{r_H}{r})^{\frac{1}{2}(3+\sqrt{1+8\gamma^2})} \, \, ,
\ee
where,
\be \label{a1}
a_1 = \frac{4 \gamma}{\; b Q_m^2(1+\sqrt{1+8\gamma^2})(3+\sqrt{1+8\gamma^2})} \phi_1 \, \, .
\ee
This correction however, vanishes at the horizon faster than $(1-\frac{r_H}{r})$.  Thus to this order, the solution  continues to be a double horizon black hole with vanishing surface gravity.  Asymptotically this correction runs to a constant  so the solution continues to be asymptotically flat to this order.

\vskip 0.3cm

\begin{flushleft}
\underline{{\bf Second order results}}
\end{flushleft}

At second order in perturbation theory the non-constant value of the scalar field we found at first order, plays the role of a source terms, resulting in corrections to the components of the metric.  We should also consider boundary conditions as follows. Since we are interested in extremal black hole solutions with vanishing surface gravity, we should have a horizon where $\beta(r)$ is finite and  $\alpha^2(r)$ has a `` double horizon". In other
words, $\alpha(r)=(r-r_H) \tilde\alpha(r)$ where $\tilde\alpha(r)$ is finite and non-zero at the horizon. It is useful to note, that by an appropriate  gauge choice, we can always take the horizon to be at $r=r_H$.  Plugging (\ref{aexpan}-\ref{phiexpan}) into equations (\ref{eqn1}-\ref{eqn4}), the solutions for $\alpha$ and $\beta$ corresponding to the above boundary conditions are:
\bea
\alpha_2(r) &=&  \alpha_H^2 a_2 (1-\frac{r_H}{r})^{1+\sqrt{1+8\gamma^2}} \, \, ,  \\ \label{41} \cr
\beta_2(r) &=& b_2 r (1-\frac{r_H}{r})^{-1+\sqrt{1+8\gamma^2}} \,\, ,
\eea
where, the constants $a_2$ and $b_2$ are
\bea
a_2&=&   \frac{\gamma}{bQ_m^2 \sqrt{1+8\gamma^2} (1+ \sqrt{1+8\gamma^2} )}\phi_2 \, \, ,
\cr \cr
&& + \frac{1}{\sqrt{1+8\gamma^2} (1+\sqrt{1+8\gamma^2})}\bigg( 2(1-\sqrt{1+8\gamma^2})(2+\sqrt{1+8\gamma^2}) \cr\crcr && \hspace{53mm} + \; 
\frac{1}{b Q_m^2}-4(1-\frac{1}{4b Q_m^2})^2 \bigg)b_2 \, \, ,\\ \label{43} \cr \cr
b_2&=& -\frac{1}{4}\; (\frac{\;1-\sqrt{1+8\gamma^2}}{\;2-\sqrt{1+8\gamma^2}} \;)  \phi_1^2 \\ \cr
\phi_2 &=& \frac{1}{4bQ_m^2} \left[ \frac{4\sqrt{1+8\gamma^2} (\sqrt{1+8\gamma^2}-1)}{\gamma (\sqrt{1+8\gamma^2}+1)(\sqrt{1+8\gamma^2}+3)} + \frac{\sqrt{1+8\gamma^2}-1}{4\gamma (2-\sqrt{1+8\gamma^2})} \right]\phi_1^2 \, \, .
\eea
These solutions, however, vanish at the horizon. Since  $\beta_1(r)$ vanishes, area of the horizon also does not change to second order in perturbation theory and is therefore independent of the asymptotic value of dilaton. Further, $\alpha_2(r)$ also vanishes at the horizon faster than $\alpha_1(r)$, thus to second order in perturbation theory, the solution continues to be a double horizon black hole with vanishing surface gravity.  Asymptotic behavior of the metric components as $r\rightarrow \infty$ is, $\alpha_2(r)\sim a_2$ and $\beta_2(r) \sim b_2 r$. Therefore, the solution continues to be asymptotically flat to this order.
\vskip 0.2cm
The scalar field also gets a correction to the second order in perturbation theory. This can be calculated in a way similar to the above analysis. We discuss this correction along with higher order corrections.

\begin{flushleft}
\underline{{\bf Higher order results}}
\end{flushleft}

We solve the system of equations $(EqA,EqB,Eq\Phi,EqC)$ order by order in the $x$-expansion.
To first order, we find that one variable, say $\phi_1$, can not be fixed by
the equations. Let us denote the value of $\phi_1$ as $K$. We thus find
$a_1$ and $b_1$ as functions of $K$.
One can check that at any order $n  \geq 2$, one can substitute the
resulting values of $(a_{m},b_{m},c_{m})$, for all $m \leq n$
from the previous orders.  Then $(EqB,Eq\Phi,EqC)$ of the
current order together with $EqA$ of order $(n-1)$,
\emph{consistently} give,
\bea\label{resultn}
b_n=b_n(K)\;\;\;;\;\;\;a_{n}=a_{n}(K)\;\;\;;\;\;\;\phi_{n}=\phi_{n}(K)\;\;\;.
\eea
as polynomials of order $n$ in terms of $K$.
\vskip 0.2cm
$K$ remains a free parameter to all orders in the $x$-expansion.
From (\ref{aexpan}), (\ref{bexpan}) and
(\ref{phiexpan}), the asymptotic values of $(a(r),b(r),\phi(r))$
are given by a sum of all the coefficients in the x-expansion of
the corresponding function. As a consequence of (\ref{resultn}),
one notices that  $(a_\infty,b_\infty,\phi_\infty)$
are free to take different values, given different choices for $K$.
The convergence of the series is not addressed in
detail, but it would be the case for small enough values for $|K|$.
\vskip 0.2cm
The value of  $\phi$ remains arbitrary at infinity, $\phi =\phi_\infty$,
while its value at the horizon is fixed to be $\phi_0$. This signifies
the presense of attractor mechanism in this theory.

\section{Conclusions}

In this paper, we studied non-supersymmetric attractor mechanism in a theory of gravity coupled
to gauge fields and scalar fields, with Born-Infeld corrections in the action.
By investigating solutions of the equations of motion, we observed the attractor behavior explicitly. 
We looked for all possible solutions which admit the criteria of being regular at the horizon 
and free in the asymptotic region. We used the perturbative approach of~\cite{0507096} to 
study the corrections to the scalar fields and took these backreaction corrections into the
metric, to show that the scalar fields are indeed drawn
to their fixed values at the horizon. 

\vskip 0.2cm

It is useful to make a few comparisons with the case of~\cite{0507096}.
In the case of Reissner-Nordstrom black holes~\cite{0507096}, there were no corrections to the
metric components to first order in perturbation theory. In the present case, $\beta(r)$ does not 
receive any correction to first order, so, the horizon area does not change to this order.
However, $\alpha(r)$ receives corrections. This, in particular means that the mass of the black hole 
starts changing from first order. Since, $\beta(r) = r$ to this order, from the $1/r$ piece of the 
$g_{rr}$ component of the metric, it is possible to extract the first order correction to mass of the BI black hole by a redefinition of the metric as (in the notation of~\cite{0507096}):
\be
M = r_H + \frac{\epsilon a_1 k r_H}{2 (1 + \epsilon a_1)}
\ee
where $a_1$ is defined in eqn. (\ref{a1}) and k is defined in eqn. (\ref{k}) and 
$\epsilon$ is a perturbation parameter.
Notice that this first order correction is positive, indicating that the lowest mass black hole
is the extremal black hole found at zeroth order. Furthermore, this correction vanishes in the
limit where the BI paramater $b \rightarrow \infty$. This is consistent with the results 
in~\cite{0507096}, where the mass of the RN black holes starts receiving correction only at
second order in perturbation theory and this should continue to hold at higher orders as well. 
To calculate the mass corrections at second and higher 
orders, one has to look at the $1/y$ term of the $g_{yy}$ component of the metric, where $y=\beta(r)$
as in~\cite{0507096}. It is in general difficult to calculate it in the present case, 
as it is not possible to find a general solution to the equation of motion 
in (\ref{ab}) due to its non-linear nature.

\vskip 0.2cm
Thus, following the analysis in~\cite{0507096} for the Reissner-Nordstrom black holes, our analysis of 
section-4 shows that the non-supersymmetric attractor behavior continues to hold in the case of 
Born-Infeld black holes as well, as long as the effective potential given in eqn. (\ref{eff}) 
has a critical point $\phi_0$ and the second derivative $\partial^2\,V_{\rm eff}$ 
evaluated at the critical point, has positive eigen values i.e., $\gamma_i > 0$. These conditions
are enough to ensure that a double zero horizon BI black hole continues to exist to all orders
in perturbation theory and thus, the attractor mechanism works to all orders in perturbation theory.
In is worth mentioning that, as in~\cite{0507096}, our perturbative analysis blows up when
$k=1/2$. This can be explicitly seen from eqn. (\ref{43}). Thus, it is expected that this
feauture continues to hold whenever $k=1/n$ for an integer $n$.
The function $\beta(r)$ does not receive any corrections to first order and further
corrections as seen in section-4 vanishes at the horizon, starting from second order eqn. (\ref{41}). 
So, the entropy also remains uncorrected in perturbation theory as for the Reisnner-Nordstrom 
black hole~\cite{0507096}. 

\vskip 0.2cm

In the case of ~\cite{0507096}, a general eight charge black
hole of heterotic string theory~\cite{0508042} can be obtained by an appropriate choice of the
scalar couplings of the gauge fields. However, in this we work, we only considered the case of
a single scalar and gauge field with the black hole carrying dyonic charges. In a more general case,
there can be further scalar couplings of the type $h_a(\phi_i){\cal L}^{a}_{BI}$, 
in the action in eqn. (\ref{last}). With such couplings and many scalar fields, 
it should be possible to study BI black
holes carrying further electric and/or magnetic charges and there could in general be multiple 
basins of attractions. In the present case, 
we explicitly showed that there are different black hole solutions
characterized by different values taken by the scalar fields of the theory at asymptotic infinity. Near the
horizon, the scalar fields however get fixed to critical values determined by the
effective potential. It should be interesting to generalize this analysis to the case of $AdS$ black holes.
Furthermore, since the Born-Infeld
terms start contributing at order $\alpha'$, together with Gauss-Bonnet terms in the string
effective action, it may also be interesting to check whether the attractor mechanism works when both sets of
terms are included.

\begin{center}
{\large {\bf Acknowledgements}}
\end{center}
Authors would like to thank Nathan Berkovits and Soo-Jong Rey for useful discussions and  comments.
BC would like to thank the FAPESP grant 05/52888-9 for financial support.
This work was supported in part by the Korea Research Foundation
Leading Scientist Grant (R02-2004-000-10150-0) and Star Faculty Grant (KRF-2005-084-C00003).


\end{document}